# AI-driven Automation as a Pre-condition for *Eudaimonia*

Anastasia Siapka

Centre for IT & IP Law, Katholieke Universiteit Leuven, Leuven, Belgium, anastasia.siapka@kuleuven.be

The debate surrounding the 'future of work' is saturated with alarmist warnings about the loss of work as an intrinsically valuable activity. Instead, the present doctoral research approaches this debate from the perspective of human flourishing (*eudaimonia*). It articulates a neo-Aristotelian interpretation according to which the prospect of mass AI-driven automation, far from being a threat, is rather desirable insofar as it facilitates humans' flourishing and, subsequently, their engagement in leisure. Drawing on virtue jurisprudence, this research further explores what this desirability may imply for the current legal order.

## 1 INTRODUCTION

Recent advances in Artificial Intelligence (AI) and robotics have rekindled fears of a workless future, with emotionally charged media narratives suggesting that AI systems and/or robots are coming to '*steal*', '*kill*' or '*destroy*' our jobs [4], [5]. The automation of work, understood as the process by which human labour is replaced by machines, is also a cause for scholarly concern across different disciplines. For some scholars, the large-scale deployment of AI in the workplace amounts to a '*Fourth Industrial Revolution*' or a '*Second Machine Age*', threatening to render human work—nay, humankind in its entirety—obsolete [3],[6]. Even despite the potential introduction of a Universal Basic Income (UBI), which could in principle guarantee citizens' livelihood, it is argued that policymakers would still need to safeguard work, since it bears intrinsic value that transcends the instrumental value of a paycheck [8]. AI-driven automation is, hence, largely framed as a threat to be counteracted by law.

Nonetheless, the axiological superiority of work as an intrinsically valuable activity and the insistence on its preservation, even if humans' sustenance could be otherwise secured, should not be taken for granted. Conversely, I argue that the prospect of automating human work through AI is, under certain conditions, desirable. To do so, I draw upon Aristotle's insights on flourishing and leisure, as these can be inferred from his *Nicomachean Ethics* and *Politics* [2]. Current normative approaches to AI-driven automation are predominantly consequentialist—assessing, for instance, its projected effects on cost-cutting in production or efficiency in service provision. Instead, I demonstrate that an approach rooted in the Aristotelian tradition could be fruitfully applied to evaluate this distinctly modern issue.

## 2 RESEARCH APPROACH

This research comprises three consecutive phases. In the **first phase** (descriptive), I have sought to define 'work'. Without clarifying work's meaning, we cannot fully understand what it is that we risk missing in the event of mass technological unemployment. Moreover, variations in the conceptualisation and evaluation of work imply corresponding variations in the perceived need as well as the measures suggested for its preservation. Therefore, with the aim of informing the normative aspects of my research, I have attempted a conceptual and axiological analysis of 'work', answering the following questions, i.e., '*what is work?*' and '*what is the value of work?*'. I have, subsequently, explored how work has been affected by technological progress over the years. Although technology has increasingly automated human tasks in the workplace over the past three centuries, most contemporary approaches to AI-driven automation focus on extrapolating to the future at the expense of the past. Ahistorical approaches, however, risk wrongfully presenting the fear of mass technological unemployment as completely novel. This is why, based on the relevant literature in computer science, engineering, the social sciences and (economic) history, I have expounded on the Industrial Revolution(s), particularly the '*Fourth Industrial Revolution*' and its distinct characteristics. Finally, I have rendered explicit the key arguments regarding the feasibility and desirability of AI-driven automation, demonstrating how they differ from the Aristotelian approach adopted in the thesis.

In the **second phase** (interpretative and theory-building), I have interpreted relevant passages from Aristotle's *Nicomachean Ethics* and *Politics*, relying on the original works alongside their secondary literature. This interpretation, in turn, has followed three steps. First, I have elucidated how Aristotle conceptualises leisure (*scholê*) and occupation (*ascholia*). His concept of leisure differs from rest, play, and entertainment, as each of these subserves one's ability for further occupation. By contrast, for Aristotle, it is occupation that should be serving leisure, not the reverse. Second, I have examined the ultimate human good in Aristotle's ethics, namely flourishing (*eudaimonia*). By explicating his conception of human flourishing as an activity of the soul in accordance with virtue/excellence (*aretê*), I have shown that leisure is indispensable to both virtue acquisition and the implied activity of the soul. However, actualising the human potential for leisure requires intentional political arrangements, which I have explored in the third step. Granting that, for Aristotle, the objective of statecraft is citizens' collective flourishing and that leisure is conducive to said flourishing, the cultivation of leisure emerges as a direct aim of politics, a shared end in which all citizens should have the realistic opportunity—although not the obligation—to partake.

In the **third and ongoing phase** (expository and normative) of my research, I flesh out the implications of this Aristotelian account of flourishing for the current legal order and, most crucially, for the debate surrounding the 'future of work'. Briefly stated, I ask: how would AI-driven automation be regulated if the *telos* (i.e., ultimate purpose or end) of law was citizens' flourishing? In responding to this question, I resort to 'virtue jurisprudence', a recently developed strand of normative legal theory that attributes primacy to the concepts of virtue/excellence and flourishing [1], [7]. Adopting a virtue-jurisprudential approach to AI-driven automation entails that—insofar as automation may generate conditions favourable to a leisure-centred polity—legislators should not only tolerate but actively incentivise AI development and adoption. Rather than seeking to preserve work by any means necessary, it is citizens' leisure that the law should be tasked with enhancing.

The remainder of the research illustrates how the law could discharge this task through its multifaceted function in society. It suggests a neo-Aristotelian interpretation, which is committed to the general structure of Aristotle's theory without necessarily subscribing to each of his doctrines. In so doing, it addresses potential objections that the suggested approach: (i) intrudes into citizens' private realm and opposes liberalism; (ii) violates state neutrality and is susceptible to abuse; (iii) hinders citizens' autonomy and freedom of choice; and (iv) is futile owing to its utopianism. The research concludes with recommendations for scholars, policymakers, AI developers, and educators.

## 3 CONCLUSION

In this way, Aristotelian ethical and political theories may enrich and expand the scope of law, making space for less conventional, even utopian for some, considerations of virtue, leisure, and flourishing. At the same time, the development of legal approaches such as virtue jurisprudence may provide concrete contexts for refining Aristotle's theories themselves and applying them to new cases of practical relevance, such as the case of AI-driven automation. Overall, this neo-Aristotelian approach not only accommodates pluralism, autonomy, and freedom of choice but further leads us to ask what the optimal conditions for flourishing—and thereby for leisure—are, conditions that legislators should seek to enable in the age of automation. This is currently an under-theorised question in the 'future of work' debate. Answering it with the help of virtue jurisprudence could yield alternative, less deterministic or dystopian, options for the pressing policy vacuum on automation and proffer novel insights into what AI promises to liberate us from and towards.

## ACKNOWLEDGMENTS

The author's research is funded by the Fonds Wetenschappelijk Onderzoek (FWO, Research Foundation – Flanders) as part of a PhD Fellowship for fundamental research (no. 1151621N/1151623N).